\documentclass{aa}

\usepackage{txfonts}
\usepackage{graphicx}
\usepackage{rotating}

\begin{document}

\title{ISO far infrared observations of the high latitude cloud L1642
       \thanks{Based on observations with ISO, an ESA project
        with instruments funded by ESA Member States (especially the PI
        countries: France, Germany, the Netherlands and the United Kingdom)
        and with the participation of ISAS and NASA} }
\subtitle{II. Correlated variations of far-infrared emissivity and 
          temperature of ``classical large'' dust particles}
\author{ K.\ Lehtinen\inst{1} \and M.\ Juvela\inst{1} \and 
         K.\ Mattila\inst{1} \and D.\ Lemke\inst{3}  \and 
         D.\ Russeil\inst{1},\inst{2}  }

\institute{Observatory, T\"ahtitorninm\"aki, P.O.
           Box 14, 00014, University of Helsinki, Finland
           \and Laboratoire d'Astrophysique de Marseille, 2 Place Le Verrier, 
                13004 Marseille, France
           \and Max-Planck-Institut f\"ur Astronomie, K\"onigstuhl 17, D-69117 
                Heidelberg, Germany}

\date{Received / accepted}

\offprints{K.\ Lehtinen (kimmo.lehtinen@helsinki.fi)}

\titlerunning{}
\authorrunning{K.\ Lehtinen et~al.} 
\abstract
{}
{Our aim is to compare the infrared properties of big,
``classical'' dust grains with visual extinction in the cloud
L1642.  In particular, we study the differences of grain
emissivity between diffuse and dense regions in the cloud.}
{The far-infrared properties of dust are based on large-scale
100\,$\mu$m and 200\,$\mu$m maps.  Extinction through the cloud
has been derived by using the star count method at $B$- and
$I$-bands, and color excess method at $J$, $H$ and $K_{\mathrm
s}$ bands. Radiative transfer calculations have been used to
study the effects of increasing absorption cross-section on the
far-infrared emission and dust temperature. } 
{Dust emissivity, measured by the ratio of far-infrared optical depth
to visual extinction, $\tau$(far-IR)/$A_{\mathrm V}$, increases with
decreasing dust temperature in L1642. There is about two-fold
increase of emissivity over the dust temperature range of
19\,K--14\,K. Radiative transfer calculations show that in order to
explain the observed decrease of dust temperature towards the centre
of L1642 an increase of absorption cross-section of dust at far-IR is
necessary.  This temperature decrease cannot be explained solely by
the attenuation of interstellar radiation field. Increased absorption
cross-section manifests itself also as an increased emissivity. We
find that, due to temperature effects, the apparent value of optical
depth $\tau_{\mathrm{app}}$(far-IR), derived from 100\,$\mu$m and
200\,$\mu$m intensities, is always lower than the true optical
depth. }
{}


\maketitle

\section{Introduction}

Recent observations have indicated that the far-IR emissivity
(measured by the ratio $\tau$(far-IR)/A$_V$) increases with
decreasing temperature (Cambr\'esy et~al.\ \cite{cambresy01}; del
Burgo et~al.\ \cite{burgo03}; Stepnik et~al.\ \cite{stepnik03};
Kramer et~al.\ \cite{kramer03}).  Probable reasons for increased
emissivity are coagulation of dust particles into large, fluffy
aggregates, and formation of ices on grain surfaces in dense
regions (Bazell \& Dwek \cite{bazell90}; Preibisch et~al.\
\cite{preibisch93}; Ossenkopf \& Henning \cite{ossenkopf94}; Dwek
\cite{dwek97a}; Cambr\'esy et~al.\ \cite{cambresy01} and
references therein).  Variations of dust optical properties are
visible even over a scale of galactic anticenter hemisphere,
anywhere where $A_{\mathrm V} \ga 1$\,mag (Cambresy et~al.\
\cite{cambresy05}).  del Burgo \& Laureijs (\cite{burgo05}) used
observations of the Taurus molecular cloud TMC-2 between
60\,$\mu$m and 200\,$\mu$m to separate the emission into a cold
and a warm component.  They found that far-IR emissivity of the
cold component is a few times larger than that of the diffuse
interstellar medium, and that the change in the properties of the
big dust particles takes place at intermediate densities of
$n$(H$_2$)$\approx 10^3$\,cm$^{-3}$.  They favoured grain-grain
coagulation over gas accretion as the reason for enhanced
emissivity because the latter process produces an increase of
emissivity by a factor of about two at most, less than that
observed by del Burgo \& Laureijs.  Also Stepnik et~al.\
(\cite{stepnik03}) suggest that grain-grain coagulation is the
reason for changes of dust properties in dense regions.

\begin{figure}
\resizebox{\hsize}{!}{\includegraphics{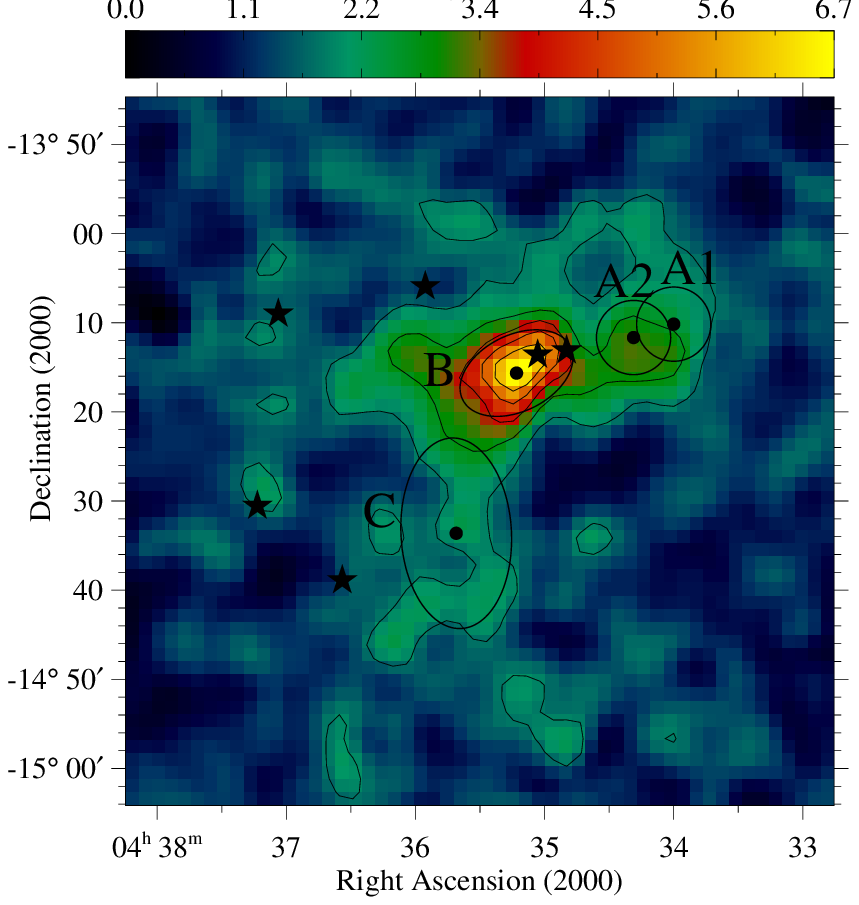}}
\caption{ Visual extinction through L1642, derived using $J$, $H$
and $K_{\mathrm s}$ data from 2MASS archive, smoothing the
individual extinctions of stars with a gaussian with
FWMH=4.5\arcmin. The IRAS sources, marked with asterisks, are
from right to left: IRAS~04325--1419, 04327--1419, 04336--1412,
04342--1444, 04347--1415 and 04349--1436.  The FWHM sizes of the
regions A1, A2, B and C as derived by Lehtinen et~al.\ (Paper~I)
from 200\,$\mu$m optical depth data are shown as ellipses and
circles.  The contours are at 1.5, 2.5, 3.5, 4.5 and 5.5\,mag }
\label{fig6}
\end{figure}

For the introduction of the high galactic latitude dark cloud L1642 we
refer to Lehtinen et~al.\ (\cite{lehtinen04}, Paper~I) and references
therein.  Paper~I was based on ISOPHOT observations in the
far-infrared, and we made the following conclusions:
\\ Based on the 200\,$\mu$m optical depth map the cloud was
   divided into three regions, called A (consists of two cloudlets,
   A1 and A2), B and C (see Fig.~\ref{fig6}).
\\ Region B, the one with the highest dust column density,
   $A_{\mathrm V} \approx 8$\,mag, is related to a temperature minimum, 
   $T_{\mathrm{d}} \approx 13.8$\,K for a $\nu^2$ emissivity.
\\ This region is close to being gravitationally bound,
   while the others are not.
\\ The lower dust temperature in region B cannot be explained
   solely by the dust attenuation of the general interstellar
   radiation field, but changes in the properties of dust particles
   are required.

In this paper we will compare the infrared properties of big
grains with visual extinction.  By comparing far-IR emission data
with optical extinction data we can study the infrared and
optical properties of dust grains.  In particular, combination of
reliable dust temperature, far-IR optical depth and visual
extinction data enables us to study the differences of grain
emissivity between diffuse and dense regions in the cloud.

\section{Data analysis}

For the data reduction of the ISO data we refer to Paper~I. 

Extinction through L1642 has been determined via star count
method at optical, and color excess method at near-IR wavelengths.
Near-IR $J$, $H$ and $K_{\mathrm{s}}$ band data have been
extracted from the 2MASS (2~Micron All Sky Survey) archive.  The
two binary T~Tauri stars near the center of the cloud have been
excluded from the 2MASS data.

\begin{figure}
\resizebox{\hsize}{!}{\includegraphics{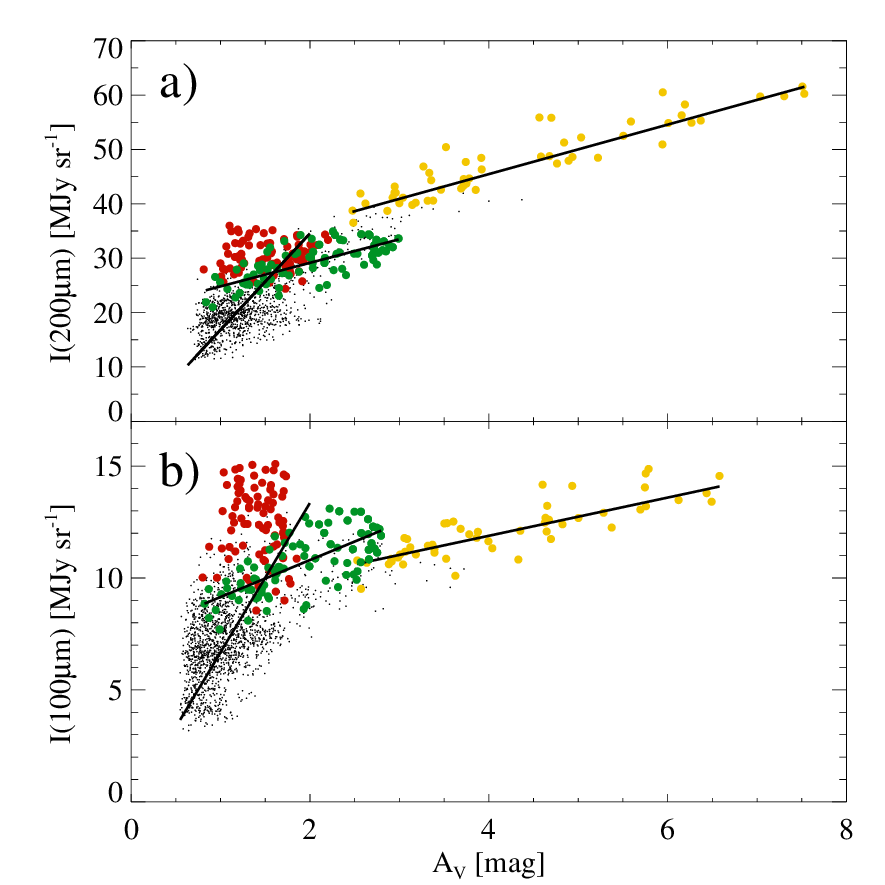}}
\caption{ {\bf a)} Relation between visual extinction and 200\,$\mu$m
surface brightness. The data points of the regions A, B and C are
marked with green, yellow and red, respectively. Linear fits
to regions A/B and interclump part of the cloud are shown. 
We have subtracted from $A_{\mathrm V}$ and
$I(200\,\mu$m) the zero-level values as fixed outside
the extent of L1642 (see Section 2.3.1).  
{\bf b)} Relation between visual extinction and 100\,$\mu$m
surface brightness. Linear fits to the regions A/B and
interclump part of the cloud are shown.  The fit of the
interclump part has been forced to go through the origin.  We
have subtracted from $A_{\mathrm V}$ and $I(100\,\mu$m) the
zero-level values as fixed outside the extent of L1642 (see
Section 2.3.1)       }
\label{fig11}
\end{figure}

\begin{figure}
\resizebox{\hsize}{!}{\includegraphics{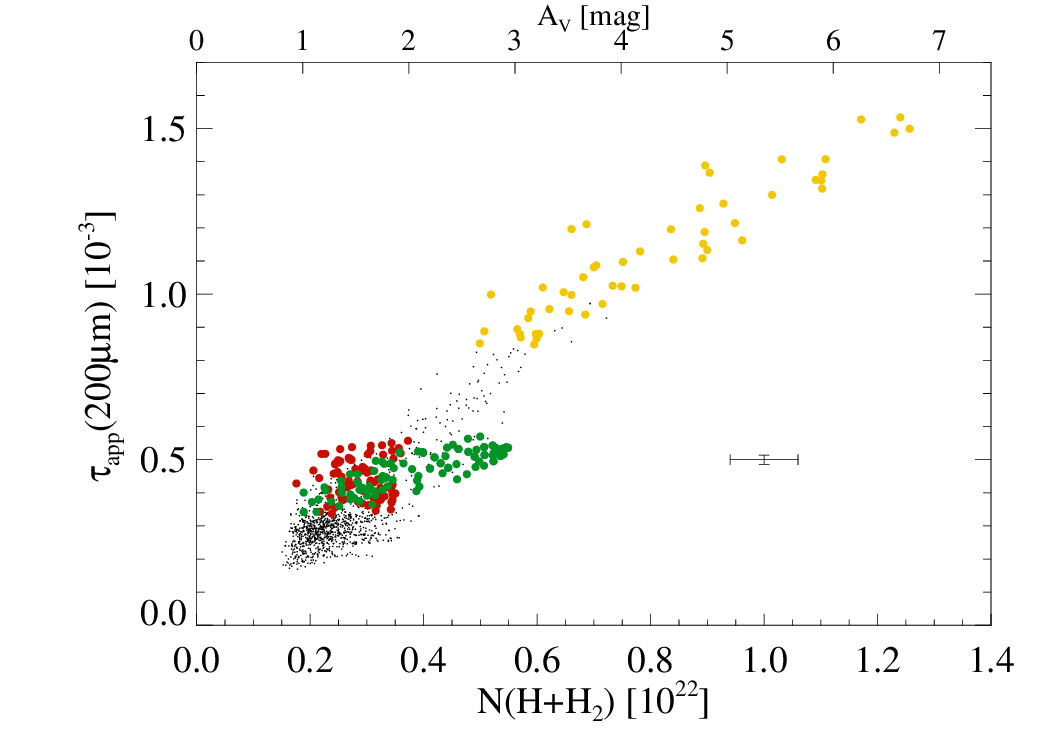}}
\caption{ Relation between apparent 200\,$\mu$m optical depth and
hydrogen column density N(H+H$_2$), derived from visual
extinction (see Sect.~3.3).  Typical pixel-to-pixel 1--$\sigma$
errors are shown. The data points of the regions A, B and C are
marked with green, yellow and red, respectively. The values of
A$_{\mathrm V}$ and 200\,$\mu$m optical depth have been derived
after subtracting from A$_{\mathrm V}$ and intensities
$I(100\,\mu$m) and $I(200\,\mu$m) the zero-level values as fixed
outside the extent of L1642 (see Section 2.3.1). }
\label{fig8}
\end{figure}

\begin{figure}
\resizebox{\hsize}{!}{\includegraphics{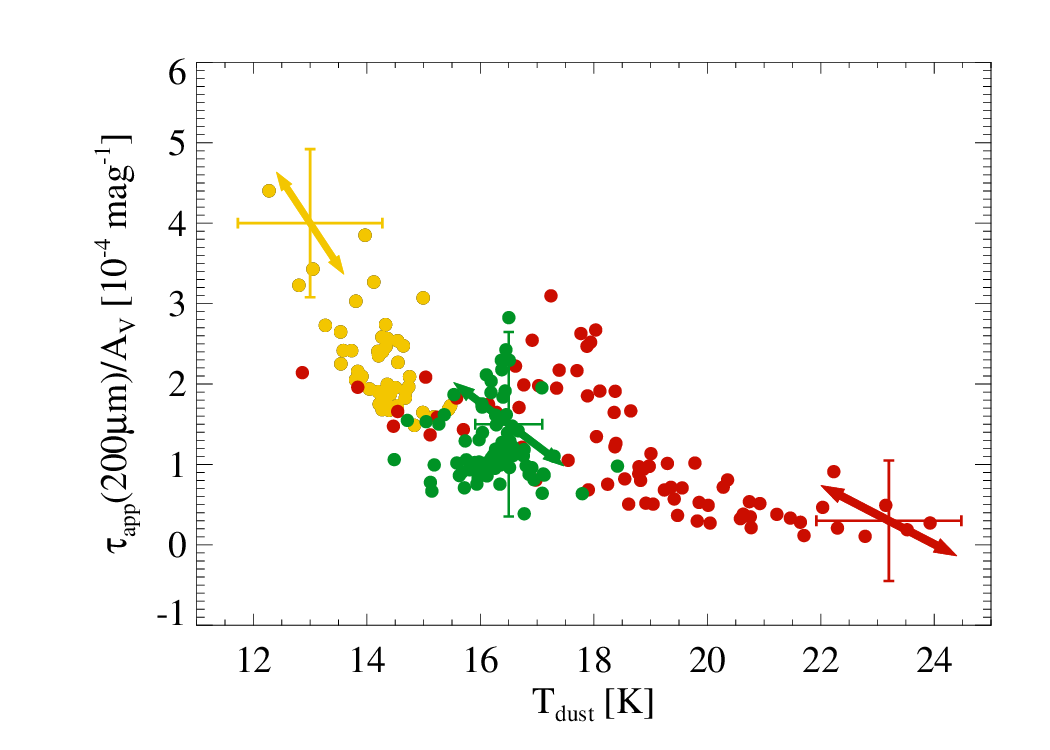}}
\caption{ The ratio $\tau_{\mathrm{app}}$(200\,$\mu$m)/A$_{\mathrm{V}}$ as
a function of dust temperature for the regions A, B and C (green,
yellow and red, respectively), as derived with the method A in
Section 3.2.  The error bars show typical pixel-to-pixel errors
within each region.  The vectors show typical direction and
magnitude (1-$\sigma$ standard deviation) of the movement of a
data point within each region due to the inverse relation between
T$_{\mathrm{dust}}$ and $\tau$(far-IR), based on Monte Carlo
calculations (see Sect.~4.2.) }
\label{fig15a}
\end{figure}

\begin{figure}
\resizebox{\hsize}{!}{\includegraphics{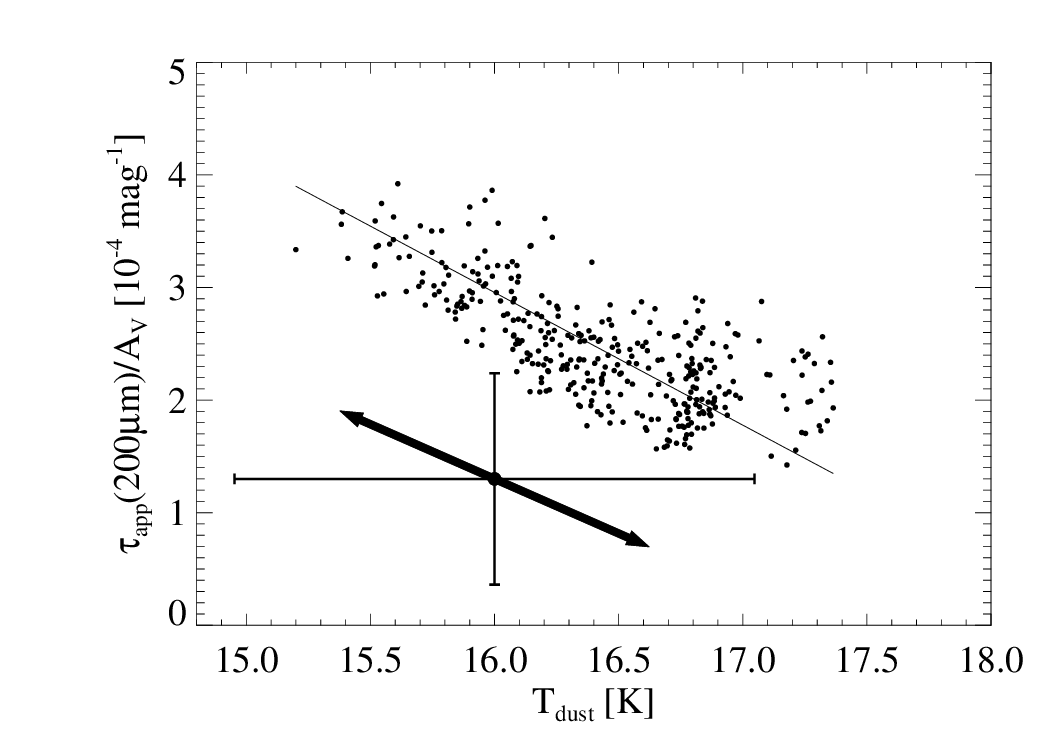}}
\caption{ The ratio $\tau_{\mathrm{app}}$(200\,$\mu$m)/A$_{\mathrm{V}}$ 
as a function of dust temperature for the
interclump part of the cloud, as derived with the method A in
Section 3.2.  To suppress noise, we have included only those
datapoints for which A$_{\mathrm{V}}>1.2$\,mag. The error bars
show typical pixel-to-pixel errors.  The straight line has a
slope of -1.2\,10$^{-4}$. The vector shows typical direction
and magnitude (1-$\sigma$ standard deviation) of the movement of
a data point due to the inverse relation between
T$_{\mathrm{dust}}$ and $\tau$(far-IR), based on Monte Carlo
calculations (see Sect.~4.2.) }
\label{fig15b}
\end{figure}

\begin{figure}
\resizebox{\hsize}{!}{\includegraphics{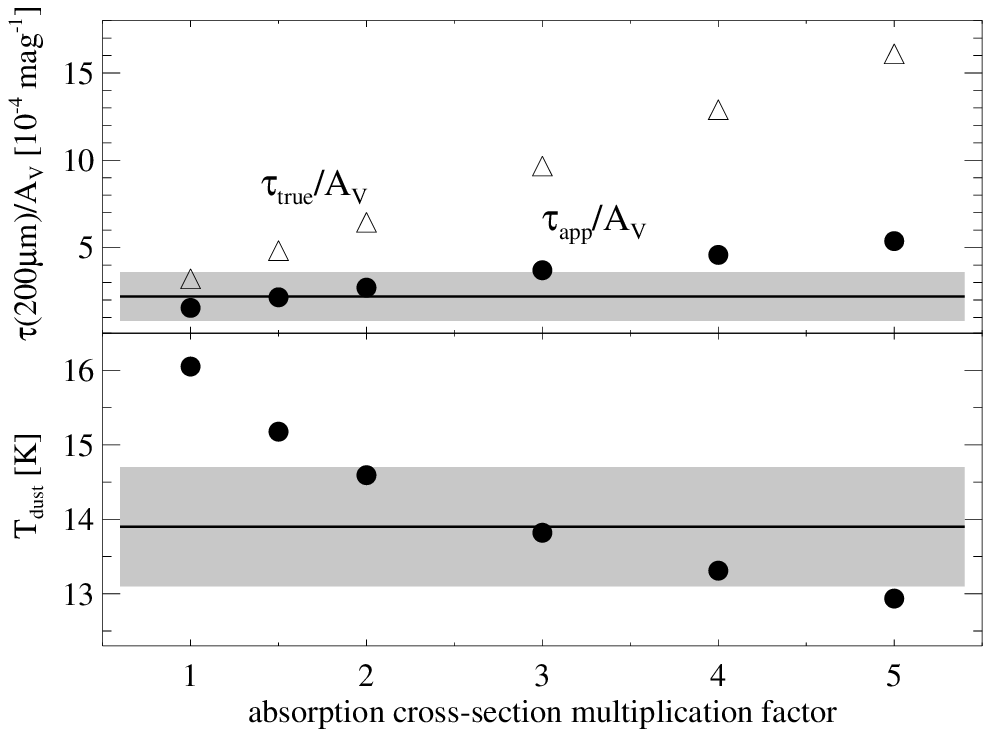}}
\caption{ Model clouds: Apparent dust temperature and emissivity
($\tau_{app}$(200\,$\mu$m)/A$_{\mathrm{V}}$) as a function of
multiplication factor $k$ for the absorption cross-section of
dust grains at far-IR (see Sect.~4.2 for details).  The dots are
values of apparent temperature and emissivity as derived from the
100\,$\mu$m and 200\,$\mu$m intensities given by the model.  The
observed mean values of temperature and emissivity in the center
of region B are marked as horizontal lines, shaded areas show the
1-$\sigma$ uncertainties. The triangles are true values of
emissivity }
\label{fig19}
\end{figure}

\begin{table*}
\caption[]{ Values of parameters in L1642, in other clouds and in
models. For a description of the different regions of L1642 see
Section~2.3.  All the data of the cloud L1642 apply to
4.5\arcmin\, resolution, except columns 4 and 6 which are for
3.5\arcmin\, resolution.  Columns 5 and 6 give the slopes of
linear fits between $I$(far-IR) and A$_{\mathrm{V}}$.  For a
description of methods A and B see Section~3.2. Columns
2--4 for L1642 give the values after the zero-levels of the
interclump dust have been subtracted (see Section 2.3).
Columns 7--10, in the case of method A, give the values after the
background levels of each region have been subtracted (see
Section 2.3). For L1642, the columns 2--4 and 7--10 (in
the case of 'method A'), give the mean values and 1-$\sigma$
standard deviations of the corresponding parameters within each
region }
\begin{flushleft}
\begin{tabular}{cccccc}
\hline\hline
 & & & & Slope of & Slope of   \\
  &  $<A_{\mathrm V}>$  &  $<I(100\,\mu$m$)>$  &  $<I(200\,\mu$m$)>$  &  
$I(100\,\mu$m$)$ vs.\ $A_{\mathrm V}$  &    
$I(200\,\mu$m$)$ vs.\ $A_{\mathrm V}$      \\
Region & [mag] & [MJy\,sr$^{-1}$] & [MJy\,sr$^{-1}$] &
[MJy\,sr$^{-1}$\,mag$^{-1}$]  &  [MJy\,sr$^{-1}$\,mag$^{-1}$]    \\
  &  &  &  &  &       \\
1 & 2 & 3 & 4 & 5 & 6      \\
\hline
L1642~A & 2.1$\pm$0.6 & 14$\pm$1 & 34$\pm$3 &  1.6$\pm$0.2
& 4.2$\pm$0.4       \\
L1642~B & 4.3$\pm$1.1 & 16$\pm$1 & 52$\pm$6 &  0.8$\pm$0.1
& 4.5$\pm$0.3        \\
L1642~C & 1.5$\pm$0.2 & 16$\pm$2 & 35$\pm$3 & - & -          \\  
L1642 interclump dust  & 1.7$\pm$0.1 & 12$\pm$2 & 28$\pm$3 &
  6.7$\pm$1.6$^{\dag}$   &  17.7$\pm$1.2                        \\   
 & & & & & \\
G~300.2--16.8 ON1 [1] & 1.8$\pm0.2$ &  & 17 &  & 9.2$\pm2.1$      \\  
G~300.2--16.8 ON2 [1] & 1.9$\pm0.2$ &  & 21 &  & 11.0$\pm2.5$     \\  
G~300.2--16.8 ON3 [1] & 2.9$\pm0.2$ &  & 35 &  & 12.1$\pm2.6$     \\  
\vspace{2cm}
 & & & & &   \\
  &  $T_{\mathrm{dust}}$  &  $<\tau(200\,\mu$m$)>$  &
$\tau(200\,\mu$m$)/A_{\mathrm V}$  &  $\sigma^H(200\mu$m)  &          \\
Region & [K]  &  [10$^{-4}$]  & [10$^{-4}$\,mag$^{-1}$] &  
[cm$^{2}$\,H-atom$^{-1}$]  &  \\
  &  &  &   & [10$^{-25}$]  &    \\
  &  7  & 8  & 9 & 10 &      \\
\hline
L1642~A (method A) & 16.3$\pm$0.6 & 1.3$\pm$0.6 & 1.3$\pm$0.5 & 
0.7$\pm$0.3  &  \\
L1642~A (method B) & 15.9$\pm$0.2 & -   & 0.8$\pm$0.1   & 0.4$\pm$0.1  &  \\
L1642~B (method A) & 14.2$\pm$0.6 & 4.9$\pm$2.5  & 2.2$\pm$0.6 & 
1.2$\pm$0.3  & \\
L1642~B (method B) & 13.4$\pm$0.3 & -   & 1.9$\pm$0.3    & 1.0$\pm$0.2  & \\
L1642~C (method A) & 18.9$\pm$2.6 & 0.7$\pm$0.5 & 1.1$\pm$0.8 & 
0.6$\pm$0.4  &      \\  
L1642 interclump dust (method A)  & 16.3$\pm$0.5  & 4.0$\pm$1.0 & 
       2.4$\pm$0.5$^{\ddag}$  &  1.3$\pm$0.3$^{\ddag}$  &  \\
L1642 interclump dust (method B)  & 17.1$\pm$1.0  & -  & 
       2.2$\pm$0.2$^{\ddag}$  &  1.2$\pm$0.1$^{\ddag}$  &  \\
 & & & & & \\
G~300.2--16.8 ON1 [1] & 17.4 & 1.9 & 1.1$\pm0.2$ &  0.6$\pm0.2$ &  \\  
G~300.2--16.8 ON2 [1] & 17.4 & 2.4 & 1.3$\pm0.2$ &  0.7$\pm0.1$ &  \\  
G~300.2--16.8 ON3 [1] & 17.5 & 4.0 & 1.4$\pm0.2$ &  0.7$\pm0.1$ & \\  
TPN [2]  &  &  & 5.3 &  2.5  &              \\  
G~301.2--16.5 [3]  &  &  &  &  0.42-0.95  &        \\  
L~183 [4]   &  &  & 3.4--3.8 &  &                   \\  
L~1780 [5]  & 15.2 &  & 2.6 &  1.4 &  \\
Diffuse ISM [6]  &  &  & 3.2 &  1.7  &              \\  
Diffuse ISM [7]  &  &  & 2.5 &  1.4  &               \\  
 & & & & \\
Model [8]  &  &  &  &  2.7  &              \\  
Model [9]   &  &  & 2.9  & 1.5  &                \\  
Model [10]  &  &  & 4.7 &  2.5  &               \\  
\hline
\end{tabular}
\newline
$^{\dag}$ The fit has been forced to go through origin \newline
$^{\ddag}$ Limited to visual extinction A$_{\mathrm V}>1.2$\,mag \newline
[1] Rawlings et~al.\ (\cite{rawlings04});
[2] Lehtinen et~al.\ (\cite{lehtinen98});
[3] Laureijs et~al.\ (\cite{laureijs96});
[4] Juvela et~al.\ (\cite{juvela02});
[5] Ridderstad et~al.\ (\cite{ridderstad06});
[6] Cambr\'esy et~al.\ (\cite{cambresy01});
[7] Dwek et~al.\ (\cite{dwek97b});
[8] Hildebrand (\cite{hildebrand83});
[9] Li \& Draine (\cite{li01b}) for diffuse ISM for $R_V$=3.1,
  interpolated from their Table~6. 
  Value of emissivity $\tau$(far-IR)/A$_{\mathrm V}$ has been
  derived from their value of $\sigma^H$(far-IR) by multiplying it with
  the ratio $N(H+H_{2})/A_{\rm V}=1.87\,10^{21}$\,cm$^{-2}$\,mag$^{-1}$ 
  (see section 3.3);
[10] D\'esert et~al.\ (\cite{desert90}) for diffuse medium.  Value
  of $\sigma^H$(far-IR) has been read from their Fig.~3 with
  about 20\% accuracy.  Value of emissivity
  $\tau$(far-IR)/A$_{\mathrm V}$ has been derived from value of
  $\sigma^H$(far-IR) by multiplying it with the ratio
  $N(H+H_{2})/A_{\rm V}=1.87\,10^{21}$\,cm$^{-2}$\,mag$^{-1}$ (see
  section 3.3)
\end{flushleft}
\label{table1}
\end{table*}

\begin{table}
\caption[]{ Error sources contributing to the pixel-to-pixel
variations of temperature, far-IR optical depth and emissivity.
See Section~2.4 for explanations }
\begin{flushleft}
\begin{tabular}{ccc}
\hline\noalign{\smallskip}
\multicolumn{2}{c}{Error source}  &  Value    \\
\noalign{\smallskip}
\hline\noalign{\smallskip}
i)    &  $<\sigma$(I(200\,$\mu$m))$>$    &  0.19\,MJy\,sr$^{-1}$   \\
ii)   &  $\sigma$($I$(100\,$\mu$m))      &  5\%                    \\  
iii)  &  $\sigma$(flatfield)             &  0.43\,MJy\,sr$^{-1}$   \\    
iv)   &  $<\sigma$(PSF)$>$               &  0.14\,MJy\,sr$^{-1}$   \\
v)    &  $<\sigma$($A_{\mathrm{V}}$)$>$          &  0.32\,mag      \\
vi)   &  $<\sigma$(zero-level(100\,$\mu$m))$>$   &  0.9\,MJy\,sr$^{-1}$   \\
vii)  &  $<\sigma$(zero-level(200\,$\mu$m))$>$   &  2.6\,MJy\,sr$^{-1}$   \\
viii) &  $<\sigma$(zero-level($A_{\mathrm{V}}$))$>$  &  0.43\,mag         \\
\noalign{\smallskip}
\hline
\end{tabular}
\end{flushleft}
\label{table2}
\end{table}

\subsection{B- and I-band extinction from starcounts}

Details of the starcounts are given in Appendix~A.  We state here
the conclusion that extinction law in L1642 is similar to that of
the diffuse interstellar matter, corresponding to 
$R_V=A_V/E_{B-V}=3.1$ (see Fig.~\ref{fig10}).

\subsection{Extinction from near-IR color excess}

In order to derive extinctions from near-IR color excesses of
stars visible through L1642 we have applied the optimized
multi-band technique of Lombardi \& Alves (\cite{lombardi01})
called NICER, which is a generalisation of the traditional color
excess method using data of two bands only.  Our data for L1642
comprise $J$, $H$ and $K_{\mathrm s}$ band magnitudes from the
2MASS archive. The surface density of stars detected at all three
bands is constant throughout the map, with an average of about
0.9 stars/$\sq\arcmin$.  The details of our extinction
determination are given in Appendix~B.  The resulting visual
extinction map, weighted with a FWHM=4.5\arcmin\, Gaussian, is
shown in Fig.~\ref{fig6}.  Region B has a maximum visual
extinction of $\sim 6.7$\,mag.  At FWHM=3\arcmin\, resolution the
maximum extinction rises to $\sim 8.2$\,mag.  Region A is seen as
a local maximum, but the angular resolution is not high enough to
resolve the clumps A1 and A2.  In general, the correspondence
between visual extinction on the one side and the 200\,$\mu$m
surface brightness and optical depth maps (see Paper~I) on the
other side is very good.  There are, however, some
differences. The 200\,$\mu$m surface brightness and optical depth
of regions A and C are about equal.  However, the maximum
extinction of region C ($\sim 2.0$\,mag) is clearly lower than
that of region A ($\sim 3.0$\,mag). Furthermore, the extinction
over region C does not have a similar regular distribution with a
single maximum as the far-IR surface brightness and optical depth
maps do.

There are differences between the USNO based and 2MASS based
extinctions. Firstly, the maximum visual extinction derived from
star counts is about half of the maximum derived from near-IR
color excesses. This can be at least partly explained by the lower
resolution and lower penetration power of the star count map.
Secondly, the relative values of extinction in regions A and C
are different; in the USNO based extinction map region C has a
larger extinction than region A, but in the 2MASS based map
the situation is reversed. The reason for this is unknown.  For
all subsequent analyses we will use the 2MASS based extinction
map.

In the following data analysis of this paper, we have used
only those datapoints for which the signal-to-noise ratio of
$A_{\mathrm{V}}$ is greater than two.  In practice, this limit
sets the lowest $A_{\mathrm{V}}$ to about 0.5\,mag.

\subsection{Zero-level of extinction and far-IR intensity}

When deriving the emissivity of dust, it is important to avoid
biases caused by improper determination of zero level of
extinction and far-IR emission.
We have determined the zero levels differently for the interclump
dust and the dense regions.

\subsubsection{Interclump dust}

We define the 'interclump dust' as the areas in the 200\,$\mu$m
map which are outside the regions A/B/C, and which have
$A_{\mathrm{V}}<2.0$\,mag.  We expect that interclump dust is low
density material located around the denser regions of L1642, and
extending beyond our 200$\mu$m map.  For interclump dust the
zero-level has to be determined at locations sufficiently far
away from L1642, outside our 200$\mu$m map. Fortunately, in the
context of another project, we have observed with ISO at
200$\mu$m several positions in the vicinity of L1642. Here we
have utilized the six positions with the lowest 200$\mu$m
intensity values, located about 2.5\degr\, South-West from the
center of L1642 (TDT (Target Dedicated Time) numbers
85500550--55).

The 100$\mu$m intensities at the same positions have been derived
from IRAS/ISSA data.  To obtain a sufficiently low noise for
these zero-level positions in the $A_{\mathrm{V}}$ maps, we used
a Gaussian with FWHM=12\arcmin\, as a smoothing function, giving
about 0.12\,mag standard deviation per map pixel.  In these six
zero-level positions, the value of
$I(100\mu$m$)=3.4\pm0.6$\,MJy\,sr$^{-1}$,
$I(200\mu$m$)=4.8\pm0.6$\,MJy\,sr$^{-1}$ and
$A_{\mathrm{V}}=0.14\pm0.11$\,mag (the given uncertainty is the
standard deviation of the parameter).

\subsubsection{Regions A, B and C}

We consider the regions A/B/C as separate clumps which are seen
in excess over the general background of L1642.  The zero-levels
of regions A/B/C have been determined in the immediate
surroundings of each of the regions. In this way, we measure
emission and extinction which is originating solely from the
region in question. Any forground/background emission by
interclump dust is in this way eliminated, at least
approximately.

For regions A and B the value of emissivity can also be derived
using the slopes of $I(100\mu$m) vs.\ $I(200\mu$m) and
$I(200\mu$m) vs.\ $A_{\mathrm{V}}$, which are independent of any
background and/or foreground zero-level subtraction. This method
will be described in Section 3.2.

\subsection{Error estimates}

In this paper our main objective is to study the {\em{variations}} of
the dust properties over the uniformly mapped L1642 cloud area.  The
different error sources contributing to the pixel-to-pixel variations
of temperature, far-IR optical depth and emissivity are listed in
Table~\ref{table2}. They are: i) the intrinsic error of each pixel of
the 200\,$\mu$m intensity map, as given by the data analysis program
PIA (ISOPHOT Interactive Analysis), ii) the pixel-to-pixel variation
of the 100\,$\mu$m IRAS map, estimated to be 5\%, iii) the error due
to the incomplete flatfield correction of the ISO map. This has been
estimated from the deviations of the intensities of adjacent pixels,
in a way similar to the actual flatfield correction estimation (see
Paper~I), iv) the error due to the non-matching point spread functions
of the 100\,$\mu$m IRAS map and the 200\,$\mu$m intensity map which
has been convolved to the IRAS resolution. This error has been
estimated by convolving the ISO map to 5.0\arcmin\, resolution instead
of the 4.5\arcmin\, resolution. As expected, the largest differences,
about 1.6\,MJy\,sr$^{-1}$, are at the positions of the largest
gradients at region B. This error source has been applied to the
200\,$\mu$m map only, v) the error of the A$_{\mathrm{V}}$ map as
given by the NICER program, vi) the error related to uncertainty
of zero-level subtraction of 100\,$\mu$m intensity (see Section~2.3),
vii) the error related to uncertainty of zero-level subtraction of
200\,$\mu$m intensity, viii) the error related to uncertainty of
zero-level subtraction of A$_{\mathrm{V}}$.  We have used the standard
deviations of the zero-level values of each region as estimates of the
last three error components.

\section{Results}

\subsection{Relation between visual extinction and far-IR surface
            brightness}

The relations between 200\,$\mu$m and 100\,$\mu$m surface
brightness and $A_{\mathrm V}$ are shown in Fig.~\ref{fig11}.
There is a clear difference between the overall relations; the
I(200\,$\mu$m) vs.\ $A_{\mathrm V}$ relation shows less scatter,
and the change of slope at $A_{\mathrm V} \approx 2-3$\,mag is
less pronounced than for I(100\,$\mu$m) vs.\ $A_{\mathrm V}$.

The parameters of linear fits to the data points I(100\,$\mu$m)
vs.\ $A_{\mathrm V}$ and I(200\,$\mu$m) vs.\ $A_{\mathrm V}$ of
different regions are given in Table~\ref{table1}. For region C,
there is no clear correlation between far-IR intensity and
$A_{\mathrm V}$, mainly because the $A_{\mathrm V}$-range in this
region is rather small (see Fig.~\ref{fig6}).  It is notable that
the highest values of I(100\,$\mu$m) are located in region C
where the visual extinction, $A_{\mathrm V} \approx 1.5$\,mag, is
lower than in regions A and B (see Fig.~\ref{fig11}b). This is due
to the higher temperature in region C (see Figs.~3 and 4 of
Paper~I). 

The general distribution of the data points is more linear at
200\,$\mu$m than at 100\,$\mu$m. This can be explained by the
fact that the temperature drop in region B influences the
100\,$\mu$m intensity more strongly than the 200\,$\mu$m
intensity because the 100\,$\mu$m intensity is on the
short-$\lambda$ side of the emission maximum.

\subsection{Determination of apparent emissivity  
            $\tau_{\mathrm{app}}$(200\,$\mu$m)/A$_V$}

When investigating the possible variations of the dust emissivity
in L1642 it is important to recognize that the optical depth as
derived from the far-infrared observations at 100\,$\mu$m and
200\,$\mu$m, i.e.\

\begin{equation}
  \tau(200\,\mu m)=I_{\nu}(200\,\mu m)/B_{\nu}(T_{\mathrm{dust}})
\end{equation}

is in general not the true optical depth of dust integrated over
the line of sight. We designate the two quantities by apparent
optical depth, $\tau_{\mathrm{app}}$(200\,$\mu$m), and true
optical depth, $\tau_{\mathrm{true}}$(200\,$\mu$m) (see Section
4.3 for a detailed discussion).

The relation between apparent 200\,$\mu$m optical depth and
$A_{\mathrm V}$ is shown in Fig.~\ref{fig8}.  At 100\,$\mu$m the
relation is exactly the same except that the
$\tau_{\mathrm{app}}$(100\,$\mu$m) values are higher by a factor of
$(200/100)^2=4$.  The Pearson correlation coefficients for the
relation are 0.85 and 0.91 for the regions A and B, respectively.

\vspace{1ex}

We have determined the apparent emissivities using two different
methods.

\vspace{1ex}

\noindent {\bf Method A.}
In this method we first subtract from the 100\,$\mu$m and 200\,$\mu$m
intensities and visual extinction the zero level values
corresponding to each region (see Section 2.3). Then the
temperature and 200\,$\mu$m optical depth are derived, on a
pixel-by-pixel basis by using Eq.~1. 
Within the different regions we derive the
following mean values of apparent emissivity: \\
region A: 1.3$\pm$0.5\,10$^{-4}$\,mag$^{-1}$    \\
region B: 2.2$\pm$0.6\,10$^{-4}$\,mag$^{-1}$    \\
region C: 1.1$\pm$0.8\,10$^{-4}$\,mag$^{-1}$    \\
interclump dust ($A_{\mathrm{V}}>$1.2\,mag): 
           2.4$\pm$0.5\,10$^{-4}$\,mag$^{-1}$   \\
The $A_{\mathrm{V}}$ values of interclump dust have been limited
to values greater than 1.2\,mag in order to suppress noise.  The
given error of emissivity is the 1-$\sigma$ standard deviation of
individual datapoints within each region.
The values of emissivity at 100\,$\mu$m are a factor of four
((200/100)$^2$) higher than the values at 200\,$\mu$m.

\vspace{1ex}

\noindent {\bf Method B.}
From the slope of the $I(200\,\mu$m) vs.\ $I(100\,\mu$m) relation
we derive the dust temperature $T_{\mathrm{dust}}$, not affected by
any constant foreground and/or background emission components. This
temperature represents a typical temperature in each region, and
is given in column 7 of Table~\ref{table1}.  
Then, the $I(200\,\mu$m) vs.\ $A_{\mathrm{V}}$ relation
(Fig.~\ref{fig11}a) gives the slope between these quantities, 
called $k$.  The apparent emissivity $\epsilon_{\mathrm{app}}$ is then
$\epsilon_{\mathrm{app}}=k/B(T_{\mathrm{dust}})$.
This method has been applied to regions A and B and interclump
dust for which we can make a reliable linear fit of
$I(200\,\mu$m) vs.\ $A_{\mathrm{V}}$.  We derive the following
mean values of apparent emissivity: \\
region A:        0.8$\pm$0.1\,10$^{-4}$\,mag$^{-1}$    \\
region B:        1.9$\pm$0.3\,10$^{-4}$\,mag$^{-1}$    \\
interclump dust: 2.2$\pm$0.2\,10$^{-4}$\,mag$^{-1}$   \\
These values are about a factor of 0.6--0.9 lower than those
derived with method A above. More importantly, both of these
independent methods give the same relative difference between
regions A, B, and interclump dust.

\subsection{Apparent absorption cross section per H-atom 
            $\sigma^H_{\mathrm{app}}(\lambda)$}

In this section we shall derive the ratio between the far-infrared
optical depth $\tau_{\mathrm{app}}(\lambda)$ and the hydrogen column density,
i.e.\ the apparent average absorption cross section per H-nucleus,
\begin{equation}
\sigma^{\mbox{H}}_{\mathrm{app}}(\lambda) = 
\frac{\tau_{\mathrm{app}}(\lambda)}{N(\mbox{H})}
\end{equation} 
We can use the near-IR extinction for estimation of the total hydrogen
column density, $N(H+H_{2})$. We adopt the value
$N(H+H_{2})/E(B-V)=5.8\,10^{21}$\,cm$^{-2}$\,mag$^{-1}$ 
valid for diffuse clouds (Bohlin et~al.\ \cite{bohlin78}) together
with
$A_{\rm V}/E(B-V)=3.1$ 
to obtain
$N(H+H_{2})/A_{\rm V}=1.87\,10^{21}$\,cm$^{-2}$\,mag$^{-1}$. 
The relation between $N(H+H_{2})$ and $\tau_{\mathrm{app}}(200\,\mu$m)
is shown in Fig.~\ref{fig8}.  Within the different regions we derive
the following mean values of $\sigma^{\mbox{H}}_{\mathrm{app}}$(200\,$\mu$m), 
calculated with the method A above (see Section~3.2): \\
region A:  0.7$\pm$0.3\,10$^{-25}$\,cm$^{2}$\,H-atom$^{-1}$  \\
region B:  1.2$\pm$0.5\,10$^{-25}$\,cm$^{2}$\,H-atom$^{-1}$  \\
region C:  0.6$\pm$0.4\,10$^{-25}$\,cm$^{2}$\,H-atom$^{-1}$  \\
interclump dust:  1.3$\pm$0.3\,10$^{-25}$\,cm$^{2}$\,H-atom$^{-1}$  \\
The given error of $\sigma^{\mbox{H}}_{\mathrm{app}}(\lambda)$ 
is 1-$\sigma$ standard deviation of individual datapoints within each
region.  The values of $\sigma^{\mbox{H}}_{\mathrm{app}}$(100\,$\mu$m)
are four times as high as the values of
$\sigma^{\mbox{H}}_{\mathrm{app}}$(200\,$\mu$m).

\subsection{ $\tau_{\mathrm{app}}$(200\,$\mu$m)/$A_V$ as a 
             function of $T_{\mathrm{dust}}$ }

Figure~\ref{fig15a} shows the emissivity
$\tau_{\mathrm{app}}$(200\,$\mu$m)/A$_V$ as a function of dust
temperature for the regions A/B/C, as derived with method A (see
Section~3.2).  The emissivity appears to increase as the temperature
decreases. Taken from Table~\ref{table2}, there is two-fold
increase of emissivity in the temperature range $\sim$19--14\,K, the
range determined by the mean temperatures of regions C and B. At the
very lowest temperature values, 12--14\,K, there is indication of a
further increase of emissivity, but the number of data points is not
sufficient to confirm that.

The emissivity as a function of dust temperature for the interclump
dust is shown in Fig.~\ref{fig15b}. The emissivity appears to
increase by a factor of about 1.5 when the dust temperature decreases
from $17$\,K to $15.5$\,K. 

In order to confirm the validity of the relation between
emissivity and dust temperature, we must exclude the possibity
that the relation is due to uncertainties in the derived dust
temperature: because the optical depth depends inversely on dust
temperature ($\tau \propto I/B_{\mathrm{dust}}$) the x- and
y-axes in Figs.~\ref{fig15a} and \ref{fig15b} are not
independent.  Any overestimation of temperature causes
underestimation of optical depth and vice versa and as a
consequence a relation like that in Figs.~\ref{fig15a} or
\ref{fig15b} may occur.  We have estimated the uncertainty of the
emissivity via Monte Carlo calculations, considering the eight
error sources as listed in Table~\ref{table2}. For details see
Appendix~C.  We conclude that the increase of emissivity with
decreasing T$_{\mathrm{dust}}$ within region B (yellow points in
Fig.~\ref{fig15a}) can be partly caused by errors in the
T$_{\mathrm{dust}}$ estimates. However, the increased emissivity
in region B relative to regions A and C cannot be explained this
way. It is found to be a real effect caused by the change of dust
properties. On the other hand, in the case of interclump
dust the range of temperature in Fig.~\ref{fig15b} is rather
small, and thus the change of emissivity is only slightly larger
than the derived uncertainty. 

\section{Discussion}

\subsection{Dust emissivity variations}

The relatively large area of 1.6$\sq \degr$ which we have mapped in
and around L1642 contains both dense and more diffuse areas with
different dust populations. The 200\,$\mu$m and 100\,$\mu$m emission,
dominated by the large particles, and the 60\,$\mu$m, 25\,$\mu$m
and 12\,$\mu$m emissions, dominated by smaller particles, peak at
separate positions in the cloud (see e.g.\ Laureijs et~al.\
\cite{laureijs87}). Compared with the situation where a sample of
distinct clouds have been studied to cover a range of different
dust properties (see e.g.\ del Burgo et~al.\ \cite{burgo03}) we
have the observational advantage that a more accurate (relative)
calibration across our single, uniformly mapped area can be
achieved. Thus, the {\it{variations}} of the dust properties over the
mapped area are more accurately determined. 

We have found in Sect.~3.4 that the emissivity of the dust,
$\tau_{\mathrm{app}}$(200\,$\mu$m)/A$_{\mathrm{V}}$, is not
constant but decreases as a function of dust temperature,
T$_{\mathrm{dust}}$.  Part of this variation is caused by the
fact that the x- and y-axes are not independent. Our error
analysis in Appendix~C has shown, however, that on the large
scale over the whole cloud area these tendencies are real.

Based on the temperature only, the emissivity of interclump dust
(T$_{\mathrm{dust}} \approx 16.3-17.1$\,K) is expected to be at about
the same level as the emissivity of region A (T$_{\mathrm{dust}}
\approx 15.9-16.3$\,K).  However, it is about two times higher, at
the same level as the emissivity in region B. This can be
explained in terms of different contributions by large grains in
different regions of L1642.  The maps of Laureijs et~al.\
(\cite{laureijs87}) show that 25\,$\mu$m emission has local maxima at
regions A and C, while 60\,$\mu$m emission has a local maximum at
region C.  This indicates that very small grains (VSGs) have a larger
abundance in regions A and C as compared to other regions.
One possible explanation for the lower far-IR emissivity of big grains
as observed in these regions A and C could be the following: the big
grains and VSGs are part of the same grain size distribution; decrease
in the number of VSGs means an increase in the number of big grains,
e.g.\ via coagulation or mantle formation. 

There is, in addition, the question whether the 100\,$\mu$m emission
in regions A and C contains a contribution from transiently heated
very small grains, which would make the derived temperature an
ill-defined quantity. This has been discussed in Appendix~D, and the
conclusion is that there is not evidence for substantial emission at
100\,$\mu$m from transiently heated VSGs in regions A or C.

\subsection{Dust emissivity versus temperature: modelling}

In Paper~I we showed that the decrease of dust temperature
towards the centre of region B in L1642 can not be explained
solely by the attenuation of the radiation field.  In the denser
parts of a cloud the dust grains are expected to stick together,
forming grains with a bigger cross-section.  We have used the
radiative transfer program of Juvela \& Padoan (\cite{juvela03})
to test if an increased value of absorption cross-section (that
is, increased emissivity) of dust grains can explain the observed
decrease of dust temperature in L1642.

The parameters of the model cloud which are based on observations
are the cloud size, radial density distribution, and visual
extinction $A_V$ at the cloud center.  Our model is tailored for
region B in L1642. The FWHM size of $\tau$(200$\mu$m) of the
model is equal to the observed one.  The density is constant up
to 0.2 times the cloud radius, outside of which it has the form
$\rho(r) \propto r^{-1.5}$. This density distribution is in
agreement with the CO observations (Russeil et~al.\
\cite{russeil02}) and the far-IR observations (Paper~I). The
cloud is heated by the Solar neighbourhood interstellar radiation
field (Mathis et~al.\ \cite{mathis83}). The grain properties used
in the model are those of Li \& Draine (\cite{li01b}).  We have
calculated a set of five models, multiplying the values of
absorption cross-sections given by Li \& Draine by a factor of
$k$=1.0, 1.5, 2.0, 3.0, 4.0, and 5.0, at all wavelengths longer
than an arbitrarily chosen threshold wavelength of 50$\mu$m.

The starting point of the model is the observed maximum visual
extinction of $A_V=6.7$\,mag at 4.5\arcmin\, resolution.
Firstly, in the case of $k$=1.0, the column density of dust is
modified until the model gives a value of $A_V$ which is equal to
the observed one; $A_V$(model) is derived by integrating the
extinction coefficient along the line of sight through the model
cloud.  Then, in subsequent calculations, with increasing values
of $k$, the true column density of dust in the model is kept the
same as in the case $k$=1.0.  The value of visual extinction
given by the model is the same for all values of $k$ because
column density of dust is constant, and the properties of dust
particles are changed only at $\lambda > 50\mu$m.

The model gives the 100$\mu$m and 200$\mu$m intensities at different
radial distances from the cloud center. The intensities are convolved
to 4.5\arcmin\, resolution.  The optical depth derived from 100$\mu$m
and 200$\mu$m intensities given by the model cloud is an apparent
optical depth, because it is obtained by mimiking the actual ISO
observations.  The true optical depth, on the other side, is derived
by integrating the absorption coefficient along the line of sight
through the model cloud.  We then derive maps of color temperature and
apparent optical depth $\tau_{\mathrm{app}}$(200\,$\mu$m).  The values
of temperature and emissivity,
$\tau_{\mathrm{app}}$(200\,$\mu$m)/$A_V$, at the centre of the model
cloud are shown in Fig.~\ref{fig19} as dots for different values of
$k$.  The observed dust temperature and emissivity at the center of
region B are marked as horizontal lines. The shaded areas show
1-$\sigma$ uncertainty of the observed values, derived by assuming
20\% absolute uncertainty for the 100$\mu$m and 200$\mu$m intensities.
Clearly, values of absorption cross-sections at far-IR enhanced by a
factor of about 2--3 are required to explain both the observed
temperature and apparent emissivity towards the center of the cloud.
The true values of emissivity, at 4.5\arcmin\, resolution, are shown
as triangles.

In the case $k=2$ the model predicts intensities at the cloud
center of about 15\,MJy\,sr$^{-1}$ and 64\,MJy\,sr$^{-1}$ at
100$\mu$m and 200$\mu$m, respectively, at 4.5\arcmin\, resolution.
These values compare well with the observed ones,
$\sim$14\,MJy\,sr$^{-1}$ and $\sim$60\,MJy\,sr$^{-1}$.

\subsection{The difference between true and apparent emissivity}

As discussed in the previous section, the apparent value of emissivity
is always smaller than its true value.  The underlying reason is that
the color temperature is biased towards too high values, leading to an
apparent optical depth which is too small.  This bias is caused by the
following facts: i) there is a temperature gradient towards the cloud
center with lower temperatures at the centre than at the edge (see
eg.\ Zucconi et~al.\ \cite{zucconi01} and Galli et~al.\
(\cite{galli02}) for a theoretical discussion), ii) graphite particles
are warmer than silicate particles which furthermore extends the
temperature range of dust particles, iii) the very small grains, which
have a larger temperature fluctuation range than the big grains, have
a noticeable contribution at 100\,$\mu$m.  The net result of the above
effects is that the emission is to some degree dominated by the warmer
dust particles; they contribute a larger fraction of the total
emission than what should be accounted for by their relative
mass. This is especially true at 100\,$\mu$m. The colour temperature
derived from 100\,$\mu$m and 200\,$\mu$m observations is therefore
larger than the proper, mass-weighted temperature.  Thus, the apparent
optical depth $\tau_{\mathrm{app}}$(200\,$\mu$m) is always smaller
than the true optical depth $\tau_{\mathrm{true}}$(200\,$\mu$m).
Consequently, the apparent emissivity is always smaller than the true
emissivity.

Irrespective of the dust temperature, an increase of absorption
cross-section of dust particles produces an increase of optical
depth by the same amount.  In Fig.~\ref{fig19}, the apparent
emissivity is seen to increase only by a factor of about 3.5 when $k$
increases from one to five. This is due to the fact that the
above mentioned effects are more pronounced when the temperature is
lower.

\subsection{Comparison with grain models and other clouds}
 
The observed two-fold increase of emissivity in L1642 is similar to
that observed in other regions; increase at 850\,$\mu$m by a factor of
$\sim$4 over a temperature range $\sim$20\,K--12\,K in the dark cloud
IC~5146 (Kramer et~al.\ \cite{kramer03}) and by a factor of $\sim$2 in
B68 (Bianchi et~al.\ \cite{bianchi03}); increase in the FIR by a
factor of $\sim$3.4 in a dense filament in Taurus (Stepnik et~al.\
\cite{stepnik03}), and increase by a factor of $>$4 in high-latitude
clouds (del Burgo et~al.\ \cite{burgo03}). 

Table~\ref{table1} gives a compilation of some observed and
theoretical properties of dust.  The theoretical values for
$\tau(200\mu$m)/$A_V$ and $\sigma^H(200\,\mu$m) in Table~\ref{table1}
are ``true'' values in our nomenclature. Because the ``apparent''
values of $\sigma^H$(far-IR) and emissivity are always smaller than
the true values (see Sect.~4.3), a comparison with the model results
is not straightforward. In fact, one should use a certain model to
derive intensity values at the two far-IR wavelengths, 100\,$\mu$m and
200\,$\mu$m, and use them to derive the apparent optical depth and
emissivity as we have done in Section 4.2.

The value of apparent $\sigma^{\mbox{H}}_{\mathrm{app}}(200\mu$m) of
region B is slightly lower than the true value obtained for diffuse
ISM by Li \& Draine (\cite{li01a}) for their carbonaceous-silicate
grain model in the case $R_{\mathrm{V}}$=3.1.  By using
Fig.~\ref{fig19} we can transform the value of
$\sigma^{\mbox{H}}_{\mathrm{app}}(200\,\mu$m) of region B into its
true value.  By assuming that the multiplication factor for emissivity
$k=2.5$, the ratio of true emissivity to apparent emissivity for the
model is about 2.5.  The value of
$\sigma^{\mbox{H}}_{\mathrm{true}}(200\mu$m) of region B would then be
about $3.0\,10^{-25}$. This is higher than any value for diffuse ISM
in Table~\ref{table1}.

The regions A and C have $<A_{\mathrm{V}}>$=1.5--2.1\,mag, slightly
higher than the extinction through typical diffuse dust.  In these
regions the value of apparent
$\sigma^{\mbox{H}}_{\mathrm{app}}(200\,\mu$m) (as measured with method
A) is clearly less than the true value obtained for diffuse ISM by Li
\& Draine.  However, we can not compare the estimated values of true
$\sigma^{\mbox{H}}_{\mathrm{true}}(200\,\mu$m) of regions A and C to
the true value of the model by Li \& Draine because we lack radiative
transfer calculations for regions A and C.
 
Mantle accretion on dust grains can increase emissivity by a factor of
about two at most (Preibisch et~al.\ \cite{preibisch93}), while
grain-grain coagulation (Bazell \& Dwek \cite{bazell90}) can increase
the emissivity by a larger amount. Thus, either of these models can
explain the observed increase of emissivity by a factor of two between
regions B and C.

\section{Conclusions}

Based on a large scale 200\,$\mu$m ISOPHOT mapping of the cloud
Lynds~1642, combined with 100\,$\mu$m IRAS/ISSA map, and an
extinction map we draw the following main conclusions:

\begin{itemize}

\item We have found that one has to make a distinction between
      apparent values of parameters such as temperature, optical
      depth and emissivity, and their true values. The apparent
      values are those which have been derived from the intensity
      maps at 100\,$\mu$m and 200\,$\mu$m. We find that the
      apparent optical depth and emissivity are always smaller
      than their true values because the apparent temperature is
      biased towards larger values

\item The apparent emissivity
      $\tau_{\mathrm{app}}$(far-IR)/A$_{\mathrm{V}}$ increases with
      decreasing dust temperature. There is about two-fold increase
      of apparent emissivity when dust temperature decreases from
      19\,K to 14\,K 

\item Radiative transfer modelling shows that an increase of
      absorption cross section of dust at far-IR is capable of
      explaining two phenomena observed in L1642: i)
      decrease of dust temperature towards the centre of the
      dense region B by an amount that is more than what can be
      explained solely by the attenuation of interstellar
      radiation field, ii) increase of apparent emissivity
      towards the colder regions. We find that an increase of the
      far-IR cross sections by a factor of about 2--3 is required
      to explain both the temperature drop and the observed
      emissivity increase in region B

\end{itemize}

\begin{acknowledgements}
The work of K.L., M.J.\ and K.M.\ has been supported by the
Finnish Academy through grants Nos.\ 1204415, 1210518 and
1201269, which is gratefully acknowledged.  This publication
makes use of data products from the Two Micron All Sky Survey,
which is a joint project of the University of Massachusetts and
the Infrared Processing and Analysis Center/California Institute
of Technology, funded by the National Aeronautics and Space
Administration and the National Science Foundation.  This
research has made use of the USNOFS Image and Catalogue Archive
operated by the United States Naval Observatory, Flagstaff
Station (http://www.nofs.navy.mil/data/fchpix/).
\end{acknowledgements}

\appendix

\begin{figure}
\resizebox{\hsize}{!}{\includegraphics{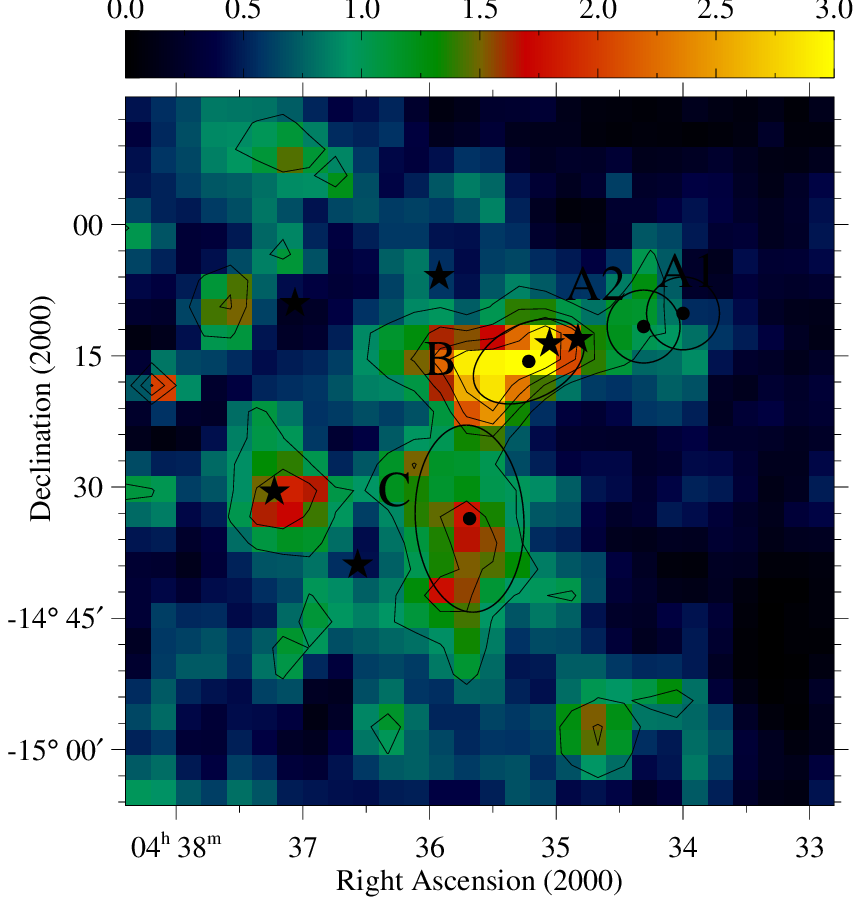}}
\caption{ Extinction through L1642 in $B$-band. 
The IRAS sources, marked with asterisks, are from right to left:
IRAS~04325--1419, 04327--1419, 04336--1412, 04342--1444, 04347--1415
and 04349--1436.  The FWHM sizes of the regions A1, A2, B and C
as derived by Lehtinen et~al.\ (Paper~I) from 200\,$\mu$m optical
depth data are shown as ellipses and circles. The contours are at
1.0, 1.5, 2.0 and 2.5\,mag }
\label{fig6a}
\end{figure}

\begin{figure}
\resizebox{\hsize}{!}{\includegraphics{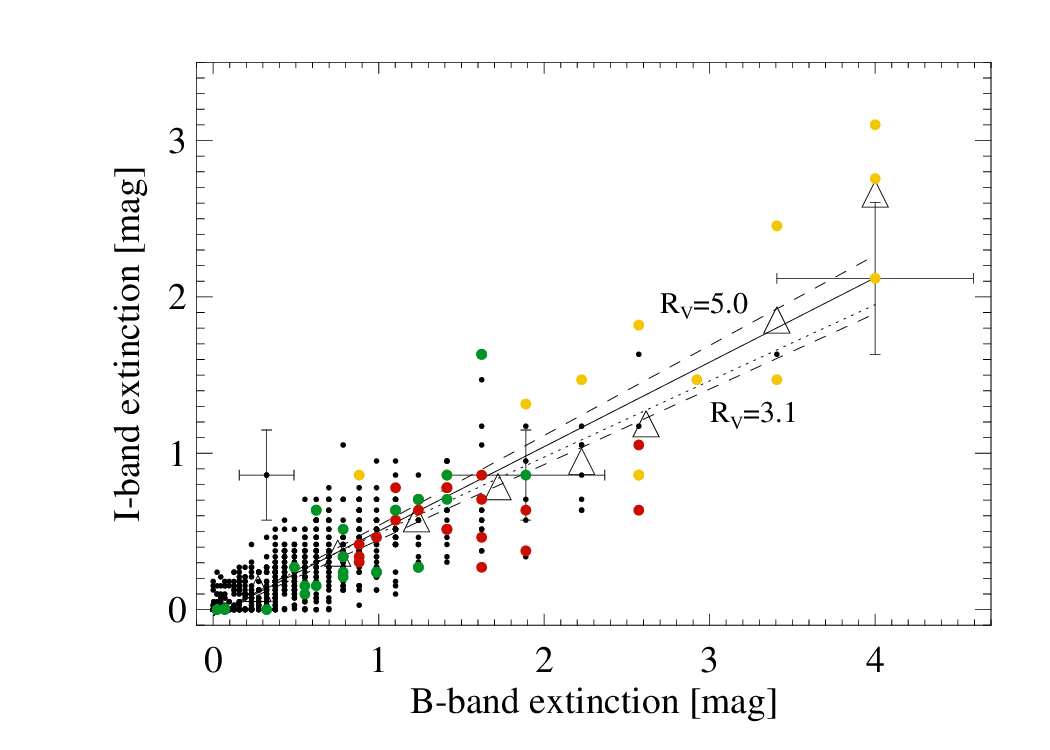}}
\caption{ Relation between star count based $B$- and $I$-band
extinction.  The data points of region A, B and C are plotted
in green, yellow and red, respectively.  The slopes of the dashed
lines correspond to the ratios of total to selective extinction
$R_V=3.1$ and $R_V=5.0$.  The solid line is a linear fit for
$A_{\mathrm B}<3.5$\,mag.  The triangles are mean values of
extinctions in 0.5\,mag $B$-band intervals up to 4.0\,mag, and the
dotted line is a linear fit of type $y=bx$ to these mean values
for $A_{\mathrm B}<3.5$\,mag.  Some examples of 1--$\sigma$
errors are shown}
\label{fig10}
\end{figure}

\begin{figure}
\resizebox{\hsize}{!}{\includegraphics{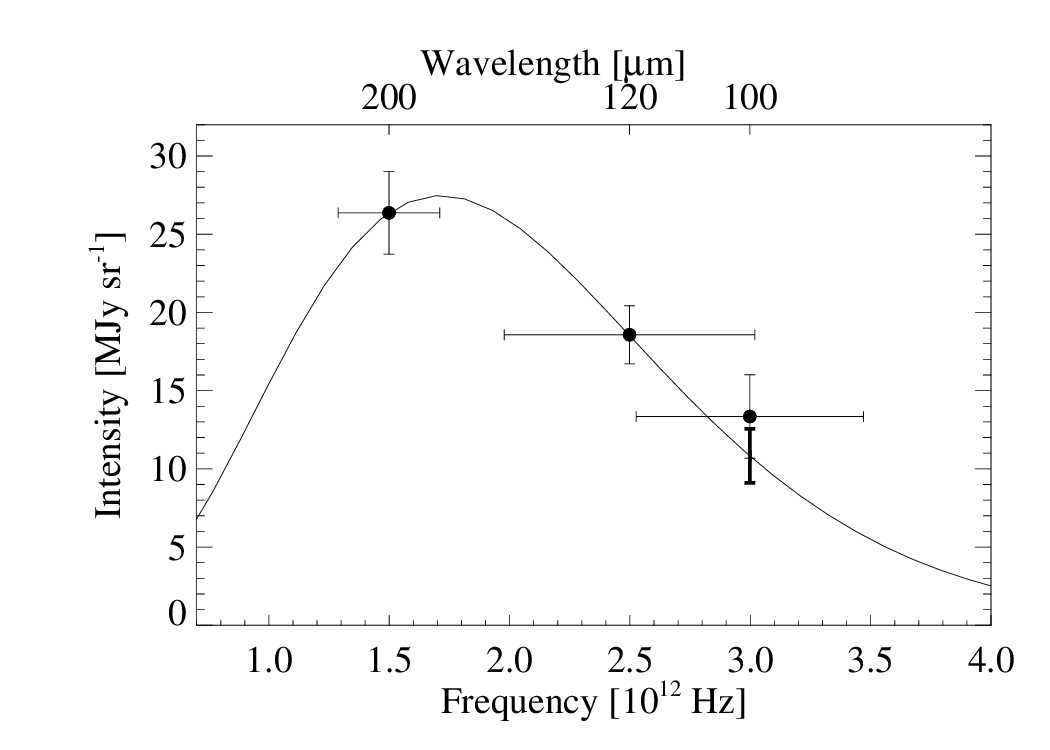}}
\caption{ The intensities measured at a position inside region C,
at 100\,$\mu$m, 120\,$\mu$m and 200\,$\mu$m.  The solid curve is a
modified black-body with a $\nu^2$ emissivity, fitted to the
120\,$\mu$m and 200\,$\mu$m data points.  The thick vertical errors at
100\,$\mu$m show the 1-$\sigma$ uncertainty of the fitted black-body,
assuming 10\% error for the 120\,$\mu$m and 200\,$\mu$m intensities.
The thin vertical bars show the estimated errors of the observed
intensities, 10\% for the ISO data at 120\,$\mu$m and 200\,$\mu$m, and
20\% for the IRAS data at 100\,$\mu$m.  The horizontal bars show the
bandwidth of each filter } 
\label{fig33}
\end{figure}

\section{ $B$- and $I$-band extinction from starcounts }

Data for star counting at $B$- and $I$-bands have been extracted from the
USNO-B1.0 archive (Monet et~al.\ \cite{monet03}).  We have used data
from the second epoch SERC-J ($B_{\mathrm J}$-band) and SERC-I
($I$-band) surveys.  For the effective wavelengths of the $B$ and $I$ bands
we have used 470\,nm and 810\,nm, respectively (see Evans \cite{evans89}). 

We were forced to use a reference area relatively close to the
L1642 (at R.A.(J2000)=4$^{\mathrm h}$35$^{\mathrm m}$00$^{\mathrm
s}$, Dec.(J2000)=-16\degr00\arcmin00\arcsec, $l$=212.90\degr,
$b$=-37.26\degr, a circular region of 1\degr radius) in order to
have the reference stars on the same plate as most of the field
stars, and thus with the same magnitude calibration. The value of
$E(B-V)$ at the reference position is about 0.12\,mag according
to the dust map of Schlegel et~al.\ (\cite{schlegel98}), which
corresponds to $A_{\mathrm V} \approx 0.4$\,mag.  Star counting
was performed in circles having a radius of 3\arcmin.
Extinctions were derived from the $\log N(m)$ vs.\ $m$ curves in
the cloud and reference areas using the method of Wolf
(\cite{wolf23}).  The pixel size of the extinction map was chosen
to be 3\arcmin.

Figure~\ref{fig6a} shows the $B$-band extinction map, after
smoothing the original map with a Gaussian with FWHM=4.5\arcmin.
The maximum extinction in the smoothed map,
$A(B)\approx$3.0\,mag, occurs close to the center of region B. In the
center of region C the extinction is about 1.8\,mag, while in
regions A1 and A2 the extinctions are about 0.8\,mag and 1.0\,mag,
respectively. 

The relation between the $B$- and $I$-band extinctions is shown in
Fig.~\ref{fig10}.  The extinction ratios at $B$- and $I$-bands
(470\,nm and 810\,nm) for the cases $R_V=3.1$ (diffuse dust) and
$R_V=5.0$ (``outer-cloud'' dust) have been adopted from Table~1 of
Mathis (\cite{mathis90}), and are shown in Fig.~\ref{fig10} as
dashed lines. These values of $R_V$ correspond to slopes of 0.48
and 0.58 in Fig.~\ref{fig10}.  The solid line is a fit of type
$y=a+bx$ for all the data points which have $A_{\mathrm B}<3.5$\,mag,
weighting by the error of extinction, giving a slope 0.54$\pm 0.02$.

The triangles are mean values of extinctions within 0.5\,mag
$B$-band intervals from $A_{\mathrm B}$=0.0--0.5\,mag up to
3.5--4.0\,mag, and the dotted line is a linear fit of type $y=bx$
to these mean values for $A_{\mathrm B}<3.5$\,mag. The derived
slope is 0.49$\pm 0.02$.  We may thus conclude that the
extinction law in L1642 is similar to that of diffuse
interstellar matter.

There is indication in Fig.~\ref{fig10} that region C has a lower
value of $R_V$ than region B. In addition, at $A_{\mathrm B} \ga
3$\,mag the relation indicates a higher value for $R_V$ inside
region B.  The number and accuracy of data points is not
sufficient for firm conclusions, though.

\section{Extinction from near-IR color excesses}

As reference fields, representing areas with ``no extinction'', we
have selected two 40\arcmin\, diameter circular areas located closely
at the same galactic latitude as L1642, at 
R.A.(J2000)=4$^{\mathrm h}$32$^{\mathrm m}$6$^{\mathrm s}$,
Dec.(J2000)=-12\degr50\arcmin27\arcsec, and 
R.A.(J2000)=4$^{\mathrm h}$41$^{\mathrm m}$22$^{\mathrm s}$,
Dec.(J2000)=-16\degr41\arcmin44\arcsec.

The observed colors of stars in the reference areas have the mean
values and standard deviations of $(J-H)$=0.52$\pm$0.17 and
$(H-K)$=0.19$\pm$0.19. They will be used as estimates of the
intrinsic colors $(J-H)_0$, $(H-K)_0$ and their variances.  For
the ratio of visual extinction to color excess we have used
$A_{\mathrm V}/E(J-H)=8.86$ and $A_{\mathrm V}/E(H-K_s)=15.98$,
which correspond to $R_V=3.1$ (Mathis \cite{mathis90}).

The pixel size in the extinction map was chosen to be the same as
the pixel size of ISO C200 detector, i.e.\ 1.5\arcmin. The
extinction value for each map pixel was derived by using a
Gaussian as a weighting function for the extinction values of
individual stars, and applying the sigma-clipping smoothing
technique of Lombardi \& Alves (\cite{lombardi01}).  The size of
the Gaussian function depends on whether we compare visual
extinction with 200$\mu$m ISO map or with maps from IRAS data. In
the case of ISO map we use a Gaussian with FWHM=3.5\arcmin,
limited by the necessity of having a sufficient signal-to-noise
ratio for the extinction. The point-spread profile of ISO at
200$\mu$m can be described by a Gaussian function having
FWHM=86\arcsec. Thus, the ISO map has been convolved with a
Gaussian having FWHM$=\sqrt(210\arcsec^2-86\arcsec^2)=192\arcsec$
when comparing with the visual extinction map having
FWHM=3.5\arcmin\, resolution.

The temperature and optical depth maps have been made by
utilizing IRAS data which have a FWHM=4.5\arcmin\, resolution,
measured from intensity profiles of point sources in the IRAS
map.  For these maps the 200$\mu$m ISO data have been convolved
with a Gaussian having
FWHM$=\sqrt(270\arcsec^2-86\arcsec^2)=256\arcsec$.  Consequently,
we use the visual extinction map with FWHM=4.5\arcmin\,
resolution in the context of temperature and optical depth maps.

There are two sources of error for the extinction map;
variance of the intrinsic colors $(J-H)_0$ and $(H-K)_0$, and
variance of the observed magnitudes of the programme stars.  In
the $A_{\mathrm V}$ map made with a FWHM=4.5\arcmin\, Gaussian
the former source dominates with a $\sim 0.29$\,mag 1-$\sigma$
error per pixel, while the latter source contributes typically
$\sim 0.08$\,mag 1-$\sigma$ error per pixel, giving a typical
total error of $\sim 0.32$\,mag per pixel.

\section{Error analysis of the emissivity}

We have estimated the uncertainty of the 200\,$\mu$m emissivity via
Monte Carlo calculations, considering the eight error sources as
listed in Table~\ref{table2}. For the emissivity estimation, the
largest error source is the error of A$_{\mathrm V}$ because it is a
large fraction of the extinction ($\sim$10--75\% for
A$_V$=6--1\,mag). On the other hand, the error of A$_V$ does not cause
any inverse emissivity vs.\ T$_{\mathrm{dust}}$ relation.  Typical
errors of emissivity and temperature, based on Monte Carlo
calculations, are plotted in Fig.~\ref{fig15a} for each of the region
A/B/C, and for interclump dust in Fig.~\ref{fig15b}.

We have used the Monte Carlo method to study into which direction
and how much a datapoint moves in a T$_{\mathrm{dust}}$ vs.\
$\tau$(far-IR) plot due to pixel-to-pixel errors in the
100\,$\mu$m and 200\,$\mu$m intensity maps. Let us first consider
the non-realistic case where only one of the intensity maps has
errors. In this case there is always a negative correlation
between T$_{\mathrm{dust}}$ and $\tau$(far-IR), just as expected.
In the case where both of the intensity maps have errors, also
positive correlation between T$_{\mathrm{dust}}$ and
$\tau$(far-IR) is possible. This happens when 100\,$\mu$m and
200\,$\mu$m intensities either increase or decrease
simultaneously.  For example, if both of them increase by a
suitable amount, the derived temperature may increase just
slightly, and then the optical depth increases too because the
relative increase of intensity may be greater than the influence
of the increase of temperature. However, a positive correlation
between T$_{\mathrm{dust}}$ and $\tau$(far-IR) happens only when
the change of temperature is minuscule. When taking into
account all the errors of Table~\ref{table2}, we find that the
slope of the inverse relation between T$_{\mathrm{dust}}$ and
emissivity has a value of $-0.5\pm$0.2\,10$^{-4}$,
$-1.1\pm$0.3\,10$^{-4}$ and $-0.4\pm$0.3\,10$^{-4}$ in region A,
B and C, respectively.  Thus, the value of the slope increases as
the temperature decreases.  For interclump dust the predicted
slope has a value of $-1.0\pm0.4\,10^{-4}$, close to the observed
value of $-1.2\,10^{-4}$. 

The vectors in Figs.~\ref{fig15a} and \ref{fig15b} show the
direction and magnitude (1-$\sigma$ standard deviation) of the
movement of datapoints caused by the inverse relation between
T$_{\mathrm{dust}}$ and emissivity, when taking all the error
sources of Table~\ref{table2} into accout.

\section{ Existence of emission from non-equilibrium very small grains }

Because the abundance of very small grains (VSGs) is enhanced in
regions A and C, we have to investigate whether there is any
significant emission from these transiently heated grains at
100\,$\mu$m.  We assume that 100\,$\mu$m is the longest wavelength
which may be affected by non-equilibrium grains. This is supported by
the observations of L1780 (Ridderstad et~al.\ \cite{ridderstad06}),
which is a high latitude cloud similar to L1642, and the
translucent/cirrus cloud as studied by Laureijs et~al.\
(\cite{laureijs00}).  In both these clouds, the intensities at
100\,$\mu$m, 120\,$\mu$m, 150\,$\mu$m and 200\,$\mu$m are all well
fitted with a single black-body function, even in the positions where
12\,$\mu$m and 60\,$\mu$m emission have their maximum values.  On the
other hand, at 80\,$\mu$m there is a significant excess over the
black-body emission from thermal equilibrium grains.

In Fig.~\ref{fig33} we plot 120\,$\mu$m and 200\,$\mu$m intensities at
the position in region C, together with a modified black-body function
fitted to them, and an estimated error of the black-body function at
100\,$\mu$m. When taking the errors into account, there is no evidence
for an excess emission at 100\,$\mu$m.  We believe that any excess
emission in region A is even smaller, because region A is colder than
region C.

The 120\,$\mu$m and 200\,$\mu$m points in Fig.~\ref{fig33} are from
ISO data, the 100\,$\mu$m point is from IRAS data (scaled to DIRBE
calibration as in Lehtinen et~al.\ \cite{lehtinen04}). 
The ON-source position in region C is at coordinates
R.A.(J2000)=4$^{\mathrm h}$36$^{\mathrm m}$7.2$^{\mathrm s}$,
Dec.(J2000)=-14\degr36\arcmin3\arcsec. 
From the ON-source (ISO TDT numbers 83901895, 83901887) intensities 
we have subtracted the mean values of intensities at two reference
positions, located at coordinates
R.A.(J2000)=4$^{\mathrm h}$33$^{\mathrm m}$17.3$^{\mathrm s}$,
Dec.(J2000)=-15\degr58\arcmin12\arcsec 
(ISO TDT numbers 83901881, 83901882, 83901890, 83901889, 83901885, 
83901893) and
R.A.(J2000)=4$^{\mathrm h}$28$^{\mathrm m}$59.4$^{\mathrm s}$,
Dec.(J2000)=-15\degr00\arcmin11\arcsec 
(ISO TDT numbers 83901896, 83901888).

\end{document}